\documentclass[fleqn]{aa501} 
\usepackage{epsf}
\usepackage{graphicx}

\newcommand{\msun}{$M_{\odot}$}

\newcommand{\mwd}{$M_{\rm wd}$}

\newcommand{\rwd}{$R_{\rm wd}$}

\newcommand{\ti}{$T_{\rm i}$}
\newcommand{\te}{$T_{\rm e}$}
\newcommand{\tm}{$T_{\rm max}$}
\newcommand{\tmbo}{$T_{\rm max,bomb}$}
\newcommand{\tmbr}{$T_{\rm max,brems}$}

\newcommand{\tsh}{$T_{\rm i,s}$}
\newcommand{\tes}{$T_{\rm e,s}$}
\newcommand{\tis}{$T_{\rm i,s}$}

\newcommand{\munit}{${\rm m}_{\rm u}$}

\newcommand{\gcs}{g\,cm$^{-2}$s$^{-1}$}
\newcommand{\gc}{g\,cm$^{-2}$}

\newcommand{\hsh}{$h_{\rm sh}$}
\newcommand{\xs}{$x_{\rm s}$}
\newcommand{\xsbo}{$x_{\rm s,bomb}$}
\newcommand{\xsbr}{$x_{\rm s,brems}$}
\newcommand{\vo}{$\upsilon_{\rm o}$}
\newcommand{\rhoo}{$\rho_{\rm o}$}
\newcommand{\mdotb}{${\dot m}B^{-2.6}$}
\newcommand{\susi}{${\dot m}\,B_7^{-2.6}$}

\newcommand{\pssym}[1]{\includegraphics[width=2mm]{#1}}

\begin{document}
\title{Accretion physics of AM Herculis binaries, I. Results from
one-dimensional stationary radiation hydrodynamics
}
\author{A. Fischer
\thanks{Present address: Lufthansa Systems GmbH, 65451 Kelsterbach, Germany}
\and  K. Beuermann
} 
\offprints{beuermann@uni-sw.gwdg.de}
\institute{ 
Universit\"ats-Sternwarte, Geismarlandstr. 11, D-37083 G\"ottingen, Germany
}
\date{Received November 30, 2000 / Accepted April 25, 2001}
\authorrunning{A. Fischer \& K. Beuermann}
\titlerunning{Accretion physics of AM Herculis binaries I.}
\abstract{ We have solved the one-dimensional stationary two-fluid
hydrodynamic equations for post-shock flows on accreting magnetic
white dwarfs simultaneous with the fully frequency and angle-dependent
radiative transfer for cyclotron radiation and bremsstrahlung.
Magnetic field strengths $B = 10$ to 100\,MG are considered. At given
$B$, this theory relates the properties of the emission region to a
single physical parameter, the mass flow density (or accretion rate
per unit area) $\dot m$. We present the normalized temperature
profiles and fit formulae for the peak electron temperature, the
geometrical shock height, and the column density of the post-shock
flow. The results apply to pillbox-shaped emission regions. With a
first-order temperature correction they can also be used for narrower
columns provided they are not too tall.
} 

\maketitle

\section{Introduction}

The thermal structure of the accretion columns on accreting magnetic
white dwarfs can be derived analytically for a single-particle fluid
and sufficiently simple assumptions on the radiative cooling (Aizu
1973, Chevalier \& Imamura 1982, Wu et al. 1994).  For the more
general case of the optically thick frequency and angle-dependent
radiative transfer in a two-fluid plasma, the coupled hydrodynamic and
radiative transfer equations have to be solved numerically (Woelk \&
Beuermann 1996, henceforth WB96). In this paper, we present results
which are improved and expanded over those of WB96. We obtain the
temperature and density profiles for plane-parallel post-shock cooling
flows and derive fit formulae for the peak electron temperature \tm,
the column density \xs, and the geometrical shock height \hsh\ as
functions of the magnetic field strength $B$ and the mass flow density
(accretion rate per unit area) $\dot m$. For low $\dot m$ and high
$B$, we show that the shock solution merges into the non-hydrodynamic
bombardment solution for an atmosphere which is heated by a stream of
fast ions and cools by cyclotron radiation (Woelk \& Beuermann 1992,
1993, henceforth WB92, WB93).

Our treatment of radiation-hydrodynamics is one\-dimensional and
stationary. The one-dimensionality implies that our solutions are
strictly applicable only to pillbox-shaped emision regions with a
width $D \gg $ \hsh\ and a stand-off distance \hsh\ $\ll$ \rwd, where
\rwd\ is the white dwarf radius. The stationarity implies that our
solutions describe the mean properties of the shocks and that aspects
like rapid fluctuations in the mass flow density and the stability
against shock oscillations are left aside.  Shock oscillations have
been treated by a number of authors (Imamura et al. 1996, Saxton \& Wu
1999, and references therein) and generally suggest that cyclotron
cooling stabilizes the flow and bremsstrahlung cooling destabilizes
it. Observationally, optical oscillations have been found in a few
polars, while the search for hard X-ray oscillations has so far
yielded only upper limits (Larsson 1992, Wolff et al. 1999,
\mbox{Imamura et al. 2000, and references therein)}.

\section{Two-fluid radiation hydrodynamics}

\subsection{General approach}

We solve the stationary, one-dimensional, two-fluid hydrodynamic
equations simultaneous with the frequency and angle-dependent
radiative transfer, closely following the approach of WB96.
We deviate from WB96 in the treatment of the shock itself. Instead of
integrating the flow through the shock with an artificial viscosity,
we adopt the presence of a strong ion shock and start the integration
with adopted values of the ion and electron temperatures (see below).
Of course, the solution now fails to reproduce the rapid rise in ion
temperature across the shock, but otherwise the results are
practically identical except for small differences at low velocities
where the flow connects to the atmosphere of the star and large
gradients revive the viscous terms again. The set of differential
equations then reads (compare Eqs. (1) to (4), (7), and (8) of WB96)
\begin{equation}
\rho \upsilon = - \dot m \label{masscons} 
\end{equation}
\begin{equation}
\rho \upsilon \,\frac{{\rm d}\upsilon}{{\rm d}x} + \frac{{\rm d}}{{\rm
d}x}(P_{\rm i} + P_{\rm e}) = - g \label{momentum} 
\end{equation}
\begin{equation}
\rho \upsilon \,\frac{{\rm d}{\cal E_{\rm i}}}{{\rm d}x} - \upsilon\,(P_{\rm
i} + {\cal E_{\rm i}}) \frac{{\rm d}\rho}{{\rm d}x} = - \Lambda_{\rm
ei} \label{ionenergy}
\end{equation}
\begin{equation}
\rho \upsilon \,\frac{{\rm d}{\cal E_{\rm e}}}{{\rm d}x} - \upsilon\,(P_{\rm
e} + {\cal E_{\rm e}}) \frac{{\rm d}\rho} {{\rm d}x} = \Lambda_{\rm
ei} -\rho\, \frac{{\rm d}F_{\rm rad}}{{\rm d}x}
\label{electronenergy}
\end{equation}
\begin{equation}
\rho\,{\rm cos}\vartheta\frac{{\rm d}I_{\nu}(\vartheta)}{{\rm d}x} =
\left [\kappa_{\nu}(\vartheta) + \sigma \right ]I_{\nu}(\vartheta) -
\kappa_{\nu}(\vartheta) B_{\nu}(T_{\rm e}) - \sigma J_{\nu}
\label{radtransfer}
\end{equation}
\vspace*{-5mm}
\begin{equation}
F_{\rm rad} = 2\pi\,\int\limits^\infty_0\,\int\limits^{+1}_{-1}\,
\,I_{\nu}(\vartheta)\,{\rm cos}\vartheta\,{\rm d(cos\vartheta)}\,{\rm d}\nu
\label{radflux}
\end{equation}
where $\rho$ is the mass density, $\upsilon$ the velocity, $\dot m$
the mass flow density, $P_{\rm i}$ and $P_{\rm e}$ the ion and
electron pressures, ${\cal E_{\rm i}}$ and ${\cal E_{\rm e}}$ ion and
electron internal energy densities, $I_{\nu}(\vartheta)$ the specific
intensity of the radiation field at frequency $\nu$ and angle
$\vartheta$, $J_{\nu}$ the angle-averaged intensity,
$\kappa_{\nu}(\vartheta)$ the angle-dependent absorption coefficient,
$\sigma$ the Thomson scattering coefficient, $B_{\nu}(T_{\rm e})$ the
Planck function at electron temperature $T_{\rm e}$, and $F_{\rm rad}$
the total radial energy flux in the radiation field. We choose the
downstream column density $x$ as the independent variable rather than
the radial coordinate or the geometrical height $h$. Following WB96, we
neglect the effects of radiation pressure and thermal conduction.

The connecting link between the hydrodynamics (Eqs. 1 to 4) and the
radiative transfer (Eqs. 5 and 6) is $F_{\rm rad}$: the electron gas
cools by radiation and is heated by Coulomb interactions with the
ions, described by the non-relativistic electron ion energy exchange
rate $\Lambda_{\rm ei}$ (Spitzer 1956, see also WB96, their Eqs. 5 and
6). The fully angle and frequency-dependent radiative transfer
accounts for cyclotron absorption, free-free absorption, and coherent
electron scattering. Our emphasis is on the largely correct treatment
of the cyclotron spectra and we accept inaccuracies of the hard X-ray
spectra caused by the neglect of Compton scattering. This still rather
general treatment ensures that our results are relevant for a wide
range of $\dot m$ including the low-$\dot m$ regime where radiative
losses by optically thick cyclotron radiation dominate.The cyclotron
absorption coefficients used here are the added coefficients for the
ordinary and the extraordinary rays (WB92). This limitation is dropped
in Sect.~\ref{spectra}, below.

We use a Rybicki code for the LTE radiative transfer and integrate the
set of equations implicitly, using a Newton scheme to iterate between
hydrodynamics and radiation transport. For more details see WB96.  Our
solution is strictly valid only for an infinite plane parallel
layer. A first-order correction to the peak electron temperature for
emission regions of finite lateral extent $D$ (Fig.~1) is discussed in
Sects.~\ref{luminosity} and \ref{tempcorr} below.

Eq.~(2) accounts for post-shock acceleration and heating of the flow
by the constant gravity term $g =\,$G\mwd/\rwd$^2$.  Within our
one-dimensional approximation which disregards the convergence of the
polar field lines, considering the variation of gravity with radius
would not be appropriate. Our approach is, therefore, limited to stand-off
distances of the shock \hsh$ \ll $\rwd. Settling solutions with
\hsh$\,\ga\,$\rwd\ are not considered.

As in WB96, we assume that the pre-shock flow is fully ionized, but
cold. Soft X-rays will photoionize the infalling matter and create a
Str\"omgren region with a temperature typical of planetary nebulae,
but for our purposes this is cold. Heating of the pre-shock electrons
by thermal conduction may be more important. Equilibrium between
diffusion and convection defines an electron precursor with a radial
extent~ $\lambda_{\rm pre} \simeq 4\times 10^{-15}T_{\rm
e,s}^{5/2}/\dot m$~cm, where $T_{\rm e,s}$ is the electron temperature
at the shock in K (Imamura et al. 1987) and $\dot m$ is in \gcs. Near
the one-fluid limit, electron and ion shock temperatures are similar,
$T_{\rm e,s} \simeq T_{\rm i,s}$, and the precursor extends to
$\lambda_{\rm pre} \simeq 0.09\,$\hsh. In a cyclotron-dominated
plane-parallel flow, however, two effects cause the precursor to be
less important: (i) the electrons never reach the peak temperature
expected from one-fluid theory and (ii) the optically thick radiative
transfer in the plane-parallel geometry sets up a radial temperature
gradient which further depresses the electron temperature at the
shock. In this paper, we do not consider thermal conduction, neglect
the presence of the electron precursor, and opt to set \tes\,$= 0$.

At $x = 0$, we adopt the Rankine-Hugoniot jump conditions for a gas
with adiabatic index 5/3, i.e. we set the post-shock density to
4\,\rhoo, the bulk velocity to \vo/4, and the pressure to
(3/4)\rhoo\,\vo$^2$, with \rhoo\ and \vo\ the density and bulk
velocity in the pre-shock flow. With \tes\,$= 0$, the ion shock
temperature is
\begin{equation}
T_{\rm i,s} = 3\,\mu_{\rm i}{\rm m_u}/(16\,{\rm k})\,\upsilon_{\rm o}^2. 
\label{tshock}
\end{equation}
We use \vo\,=\,(2G\mwd/\rwd)$^{1/2}$ with Nauenberg's (1972) relation
between mass and radius of the white dwarf. $\mu_{\rm i}$ is the
molecular weight of the ions, \munit\ the mass unit, and
k \mbox{the Boltzmann constant}.  

All numerical calculations are performed for a hydrogen plasma with
$\mu = 0.5$ and $\mu_{\rm i} = \mu_{\rm e} = \mu_{\rm Z} =1$, where
$\mu_{\rm e}$ is the number of nucleons per electron, $\mu$ is the
molecular weight of all particles, and $\mu_{\rm Z}$ the molecular
weight of the ions weighted with $Z_{\rm k}^2$. We include the
molecular weight dependence in our equations in order to allow
conversion to other compositions, e.g., a fully ionized plasma of
solar composition with $\mu = 0.617$, $\mu_{\rm i} = 1.297$, $\mu_{\rm
e} = 1.176$, and $\mu_{\rm Z} = 0.927$.

\subsection{\label{heatcool} Bremsstrahlung and cyclotron emissivities}

The frequency-integrated volume emissivity for brems\-strahlung is 
\begin{equation}
\epsilon_{\rm brems} = {\rm c}_2 T^{1/2} n_{\rm e} 
\sum\limits_{\rm k} n_{\rm k} Z_{\rm k}^2 
= \frac{{\rm c}_2}{({\rm k}\,{\rm m}_{\rm u}^3)^{1/2}}
\frac{\mu^{1/2}}{\mu_{\rm e} \mu_{\rm Z}}\,P^{1/2}\rho^{3/2}  
\label{epsbrems}
\end{equation}
where c$_2 = 1.43\,10^{-27}$ cgs-units, $n_{\rm e}$ is the number
density of the electrons, $n_{\rm k}$ the number density of the ions
of charge $Z_{\rm k}$e,\linebreak $P$ is the gas pressure, and $\rho$
the mass density. The Thomson scattering optical depth of a
bremsstrahlung dominated flow parallel to the flow is of order unity,
implying that bremsstrahlung is essentially optically thin (Aizu
1973).
 
We use the cyclotron absorption coefficients for the ordinary and
extraordinary rays given by Chanmugam \& Dulk (1981), Thompson \&
Cawthorne (1987), and WB92. The total cyclotron emissivity of
non-relativistic electrons integrated over wavelength and solid angle
is
\begin{equation}
\epsilon_{\rm cyc}
\propto n_{\rm e}\,T\,B^2 \propto (\mu/\mu_{\rm e})\,B^2 P.
\label{epscyc}
\end{equation}
In the columns considered, cyclotron radiation is optically thin in
the higher harmonics, but is always optically thick in the first few
harmonics, and the temperature distribution in a cyclotron-dominated
emission region can properly be calculated only by solving the coupled
radiation-hydrodynamic equations.

\begin{figure}[t]
\includegraphics[width=8.8cm]{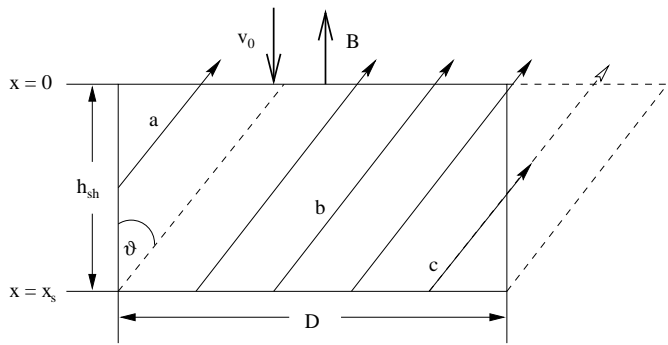}
\caption[]{\label{schematic} Schematics of the emission region. The
region is bounded at the top by the shock front and at the bottom by
the white dwarf.}
\end{figure}

\subsection{Geometry of the emission region}

Fig.~1 shows the schematic of an emission region with finite lateral
extent $D$.  The shock is located at $h =\,$\hsh\ above the white
dwarf surface. The downstream column density is $x = 0$ at the shock
and and $x\,=\,$\xs\ at the surface of the star, with $x$ and $h$
being related by d$x = -\rho$\,d$h$. The gravity vector {\bf g} and
the magnetic field vector {\bf B} are taken parallel to the flow
lines.  The radiation-hydrodynamic equations are solved for layers of
infinite $D$ to yield the run of electron temperature and mass
density, $T_{\rm e}(x)$ and $\rho(x)$. These profiles are later
employed to calculate the outgoing spectra for emission regions with
finite $D$ by ray tracing, i.e. by adding the contributions from an
appropriate number of rays (Fig.~1 and Sect.~\ref{spectra}). This
procedure is not self-consistent if optically thick radiative losses
occur from the sides of the column. An appropriate first-order
correction to the temperature structure derived for the infinite layer
is discussed in Sects.~\ref{luminosity} and \ref{tempcorr}, below.
The treatment of really tall columns requires a different approach
which specifically allows for the emission from the sides of the
column (Wu et al. 1994).

Radiation intercepted by the white dwarf is either reflected or
absorbed and reemitted by its locally heated atmosphere. We assume
coherent scattering of hard X-rays using the frequency-dependent
reflection albedo $A_{\nu}$ of van Teeseling et al. (1994). The
fraction $1 - A_{\nu}$ of the energy is re-emitted in the UV and soft
X-ray regime and is not considered in this paper. 

\subsection{Limiting cases}

Here, we consider simple limiting cases which can, in part, be solved
analytically. Below, we shall discuss our numerical results in terms
of these limiting solutions. The high $\dot m$, low $B$ limit is the
bremsstrahlung-dominated one-fluid solution. In the opposite limit of
low $\dot m$, high $B$ one enters the non-hydrodynamic regime (Lamb \&
Masters 1977). Here, the bombardment solution of a static atmosphere
heated by a stream of fast ions and cooling by cyclotron emission is
an appropriate approximation (Kuijpers \& Pringle 1982, WB92, WB93).

The one-dimensional, one-fluid hydrodynamic equations with simple
terms for optically thin cooling can be solved analytically (Aizu
1973, Chevalier \& Imamura 1982). Integration of Eq. (2) with Eq. (1),
$P = P_{\rm i} + P_{\rm e}$, and $g = 0$ yields \mbox{$P = \dot
m(\upsilon_{\rm o} - \upsilon)$} which allows to express the
emissivities of Sect.~\ref{heatcool} as \mbox{$\epsilon_{\rm brems}
\propto \dot m^2~{\rm f}(\upsilon)$} and $\epsilon_{\rm cyc} \propto
\dot m~{\rm g}(\upsilon)$, with f and g being functions of the flow
velocity $\upsilon$ and with additional dependencies on the $\mu$'s
and $B$ contained in the proportionality factors. Integration of the
energy equation over $\upsilon$ yields expressions for the column
density \xs\ and the geometrical shock height \hsh\ which reflect the
parameter dependence of $\epsilon$,
\begin{eqnarray}
x_{\rm s} & \propto & \frac{\dot m^2}{\epsilon}\hspace{3.3mm}
\left\{
\begin{array}{@{\hspace{4.7mm}}lcl@{\quad}l@{\quad}l}
x_{\rm s,brems}&=&{\rm constant}&{\rm for}& \epsilon = \epsilon_{\rm brems}\\ 
x_{\rm s,cyc}&\propto & \dot m  &{\rm for}& \epsilon = \epsilon_{\rm cyc},
\end{array}
\right.
\label{thinxs}\\
h_{\rm sh} & \propto & \frac{\dot m}{\epsilon}\hspace{5mm}
\left\{
\begin{array}{@{\hspace{3mm}}lcl@{\quad}l@{\quad}l}
h_{\rm sh,brems}&=&1/\dot m &{\rm for}& \epsilon = \epsilon_{\rm brems}\\ 
h_{\rm sh,cyc}&\propto &{\rm constant} &{\rm for}& \epsilon = 
\epsilon_{\rm cyc}. 
\end{array}
\right.
\label{thinhsh}
\end{eqnarray}
In the flows considered here, bremsstrahlung is close to optically
thin and the analytical solution is quantitatively corroborated by our
numerical results. Cyclotron emission, on the other hand, is optically
thick in the lower harmonics which reduces the effective emissivity
and inflates the emission region. While the $\dot m$-dependence of
cyclotron-dominated columns in Eqs.~(\ref{thinxs}) and (\ref{thinhsh})
is recovered in our numerical calculations, not surprisingly, the
numerically derived sizes of \xs\ and \hsh\ are much larger than those
predicted by the (unquoted) proportionality factors for cyclotron
cooling in Eqs.~(\ref{thinxs}) and (\ref{thinhsh}).

\begin{figure*}[t]
\includegraphics[width=7.2cm,angle=270]{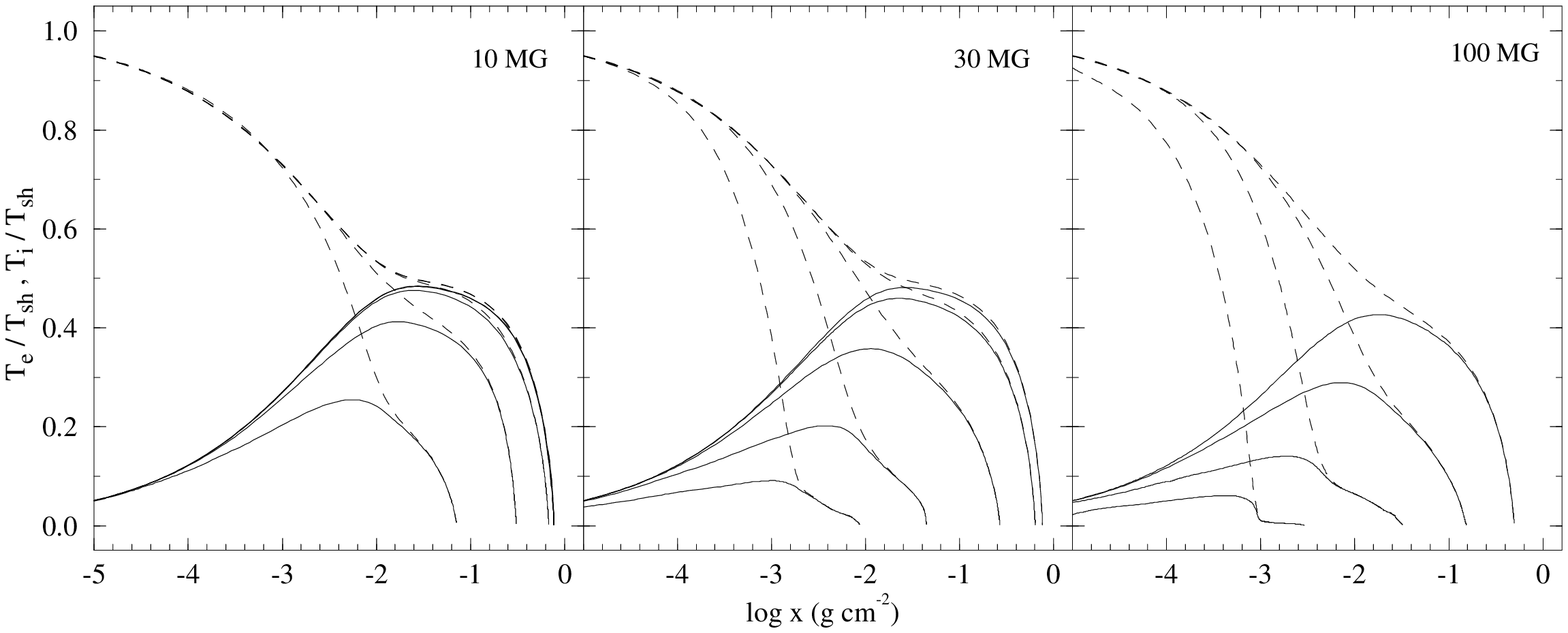}
\caption[]{\label{tempprof} Temperature profiles for the ions (dashed
curves) and electrons (solid curves) as functions of column density
$x$ for \mwd = 0.6\,\msun\ and field strengths of 10\,MG (left panel),
30\,MG (center panel) and 100\,MG (right panel). The individual curves
are for mass flow densities $\dot m = 100, 10, 1, 10^{-1}$ and
$10^{-2}\,$\gcs\ (from top). In the left panel, the curves for 100 and
10\,\gcs are indistinguishable. In the right panel, the bottom curve
is for $10^{-1}\,$\gcs. From Eqs.~(\ref{xsbrems}) and
(\ref{tmaxbrems}), $x_{\rm s, brems} = 0.783$\,\gc\ (log\,$x_{\rm s,
brems} = -0.106$) and \tmbr\ = \tsh/2.}
\end{figure*}

\subsection{\label{shocksol}Bremsstrahlung-dominated shock solution}

For a strong shock in a one-fluid plasma with adiabatic index 5/3, the
normalized post-shock velocity $\omega =4\upsilon/$\vo\ varies between
1 and 0. The column density $x$ measured from the shock is related to
$\omega $ by (Aizu 1973, Chevalier \& Imamura 1982)
\begin{equation}
x=c_1\upsilon_{\rm o}^2\left[\sqrt{3}-\frac{\pi}{3}-\frac{1+\omega }{2}\sqrt{4\omega -\omega ^2}
+{\rm cos}^{-1}\left(1-\frac{\omega }{2}\right)\right].
\label{x-w}
\end{equation}
The total column density and shock height are given by
\begin{eqnarray}
x_{\rm s, brems} & = & c_1\,\upsilon_{\rm o}^2\,(\sqrt{3}-\pi/3)
~~~{\rm g\,cm}^{-2}
\label{xsbrems}\\
h_{\rm sh, brems} & = & c_1\,\upsilon_{\rm o}^3\,\left(39\sqrt{3}-20\pi\right)/
(48\,\dot m)~~~{\rm cm},
\label{hshbrems}
\end{eqnarray}  
\noindent where $c_1 = ({\rm k}^{1/2}{\rm m}_{\rm u}^{3/2}/4\,c_2)
(\mu_{\rm e}\,\mu_{\rm Z}/\mu^{1/2})$ with values of $6.22~10^{-18}$
cgs for pure hydrogen and $6.10~10^{-18}$ cgs for solar composition. 
The temperature profile follows from pressure equilibrium $P = \rho
\upsilon (\upsilon_{\rm o} - \upsilon$), the equation of state for the
ideal gas, and Eq.~(1) as
\begin{equation}
T  =  \frac{1}{3}\left(4\omega  - \omega ^2\right)T_{\rm max,brems}
\label{tbrems}
\end{equation}
with 
\begin{equation}
T_{\rm max,brems} = 3\,\mu {\rm m_u}/(16\,{\rm k})\,\upsilon_{\rm o}^2
= (\mu/\mu_{\rm i})\,T_{\rm i,s}.
\label{tmaxbrems}
\end{equation}
Our two-fluid calculations for high $\dot m$, low $B$ reproduce the
temperature profile $T(x)$ given by Eq.~(\ref{tbrems}) with
(\ref{x-w}) and (\ref{tmaxbrems}) to better than 1\,\% of \tmbr,
except for the initial equilibration layer which is infinitely thin in
the analytic calculation and has a finite thickness with rising
electron temperature in our calculations.  As an aside, we note that
$T$/\tmbr$\,\simeq (1-x/$\xs)$^{0.59}$ with an rms error of less than
1\%.

\subsection{\label{bombsol}Cyclotron-dominated bombardment solution}

The bombardment solution involves by nature a two-fluid approach. WB92
solved this case using a Fokker-Planck formalism to calculate the
stopping length of the ions and a Feautrier code for the radiative
transfer.  WB93 (their Eqs. 8, 9) provided power law fits to their
numerical results for the column density and the peak electron 
temperature. Since the ions are slowed down by collisions with
atmospheric electrons, a factor $\mu_{\rm e}$ appears in \xs:
\begin{eqnarray}
x_{\rm s, bomb} & = & 3.94\times 10^{-2}\,\mu_{\rm e} 
(\dot m B_7^{-2.6})^{0.30} M_{\rm wd}^{1.72}~~~{\rm g\,cm}^{-2}
\label{xsbomb}\\
T_{\rm max,bomb} & = & 1.28\times 10^9\,(\dot m B_7^{-2.6})^{0.42} 
M_{\rm wd}^{0.66}~~~~~K.
\label{tmbomb}
\end{eqnarray}   
Here, $\dot m$ is in \gcs, $B_7$ is in units of $10^7$\,G, and \mwd\
is in solar masses.  These fits are very close to the quasi-analytical
expressions of Eqs. (5) and (6) of WB93\footnote{Note the misprint in
Eq. (6) of WB93 which should read $B^{0.85}$ instead of $B^{1.85}$.}. 

With increasing $\dot m$, a shock develops which is initially
cyclotron-dominated and ultimately bremsstrahlung-dominated. Since
$T_{\rm max,bomb}$ reaches \tmbr\ at some intermediate $\dot m$, we
expect a smooth transition in peak temperature between these cases.
The situation is quite different for \xs, however. At the $\dot m$
where \tmbo\ equals \tmbr, \mbox{\xsbo\ and \xsbr}\, differ by more
than two orders of magnitude.  The run of \xs($\dot m$) between these
two limiting cases can be determined only with a
radiation-hydrodynamical approach.

The bombardment solution does not predict the geometrical scale height
of the heated atmosphere which we expect to lie between that of a
corona with an external pressure $P = 0$ and that of a layer
compressed by the ram pressure $P =\,$\rhoo\,\vo$^2$.
 
\subsection{Parameterization of the results}

In the bombardment solution (Eqs. 17, 18), the dependence of \xs\ and 
\tm\ on \mdotb\ is obtained from the equilibrium between the
\mbox{energy} gain by accretion, $F_{\rm acc} \propto \dot m$, and the
energy loss by optically thick cyclotron radiation, $F_{\rm cyc}
\propto T_{\rm max}\,\omega_*^3$, where $\omega_* = m_*\,\omega_{\rm
c}$ is the high-frequency cutoff of the cyclotron spectrum and
$\omega_{\rm c} \propto B$ the cyclotron frequency. We determine the
limiting harmonic number $m_*$ from the cyclotron calculations of
Chanmugam \& Langer (1991; their Fig. 5) as \mbox{$m_* =
4.43\,\Lambda_4^a\,T_8^b$} with $a \simeq 0.12$ and $b \simeq
0.40$. This approximation is valid near depth parameters $\Lambda_4
=\Lambda/10^4 \simeq 1$ and temperatures $T_8 = T_{\rm
max}/(10^8\,{\rm K}) \simeq 1$ and is more adequate for the
cyclotron-dominated emission regions on polars than the frequently
quoted formula of Wada et al. (1980). Replacing $\omega_*$ with
$\Lambda \propto x_{\rm s}/B$ in $F_{\rm cyc}$ and equating the
accretion and radiative energy fluxes yields the result that a 
power of \tm\ is proportional to $\dot m\,B^{3(a-1)} \simeq
~$\mdotb. The same holds for \xs.

We find that the cyclotron-dominated shocks at low $\dot m$ behave
similarly to bombarded atmospheres in that their thermal properties,
too, depend on \mdotb. The individual temperature profiles $T_{\rm
e}(x)$ for different $\dot m, B$ with the same \mdotb\ coincide only
in an approximate way, but the dependency on \mdotb\ holds quite well
for the two characteristic values of each profile, \tm\ and \xs.  If we
leave the exponent $\alpha$ in $\dot m\,B^{\alpha}$ as a fit variable,
the smallest scatter in \tm\ and \xs\ as functions of $\dot
m\,B^{\alpha}$ is, in fact, obtained for $\alpha = -2.6\pm0.2$.

Our model calculations cover magnetic field strengths $B = 10-100$\,MG
and mass flow densities $\dot m = 10^{-2}-10^2$ \gcs.  In what
follows, we present first the temperature and density profiles along
the flow lines. From these, we obtain \tm, \xs, and \hsh\ as the
characteristic parameters of the post-shock flow which are presented
in an appropriate way as functions of \mdotb.

\begin{figure}[t]
\includegraphics[angle=270,width=8.5cm]{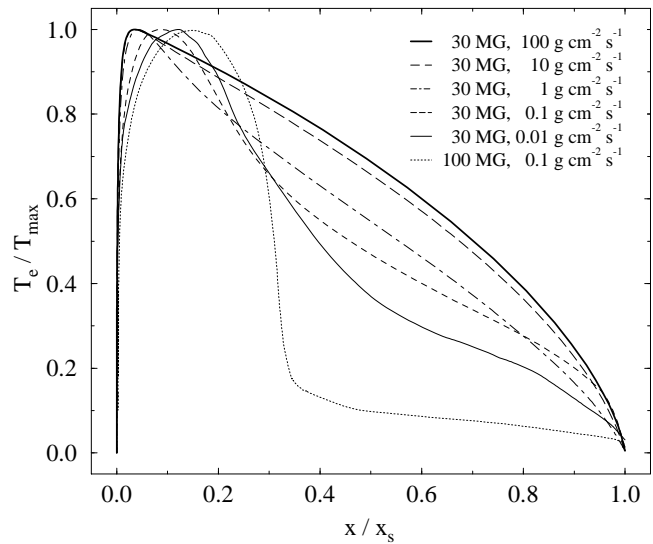}
\caption[]{\label{tnorm} Normalized electron temperature distributions for
\mwd\,= 0.6\,\msun\ and the values of $B$ and $\dot m$ given in the
figure. }
\end{figure}

\subsection{Electron temperature profiles for an infinite layer}

Fig.~\ref{tempprof} shows the temperature profiles \ti($x$) and
\te($x$) for \mwd \,= 0.6\,\msun\ and several $\dot m$-$B$
combinations on a logarithmic depth scale which emphasizes the initial
rise of the profiles. These profiles display a substantial spread in
\xs\ and in \tm, reflecting the influence of cyclotron cooling.  At
$10^{-2}$\,\gcs, 100\,MG, cyclotron cooling has reduced \tm\ to 6\%
and \xs\ to 0.3\% of the respective values for the pure bremsstrahlung
solution. We have confidence in our numerical results because they
accurately reproduce the analytic bremsstrahlung solution (see above).

Fig.~\ref{tnorm} displays the normalized profiles of the electron
temperature, \te/\tm\ vs. $x$/\xs, for different $\dot m, B$
combinations, covering the range from a bremsstrahlung-dominated flow
with 100\,\gcs, 30\,MG (fat solid curve) to $10^{-2}$\,\gcs, 100\,MG
near the non-hydrodynamic limit (dotted curce). They represent an
approximate sequence in \mdotb, but not surprisingly, the shapes
differ somewhat for different $\dot m$ and $B$ combinations
with the same value of \mdotb\ (not shown in Fig.~\ref{tnorm}).

Equilibration between electron and ion temperatures is reached at
column densities of $\sim 10^{-3} \ldots 10^{-1}$\,\gcs\ depending on
$\dot m$ and $B$ (Fig.~2). At 100\,\gcs, 10\,MG, electrons and ions
equilibrate as early as $\sim 0.03$\,\xs, while at $10^{-2}$\,\gcs,
100\,MG, equilibration length and \xs\ are of the same order,
indicating the approach to the non-hydrodynamic regime. A peculiar
feature of the latter profile is the extended low-temperature tail
which was not adequately resolved by WB96. This tail appears when
equilibration occurs near the temperature at which cyclotron cooling
becomes ineffective and the density is sufficiently high for
bremsstrahlung to take over. It is hydrodynamic in origin. Apart from
the tail, the temperature profile at $10^{-2}$\,\gcs, 100\,MG is very
close to that obtained by the non-hydrodynamic approach of WB92,
WB93. The low-temperature tail is responsible for a low-temperature
thermal emission component with k$T \la 1$\,keV.
 
\begin{figure}[t]
\includegraphics[angle=270,width=8.5cm]{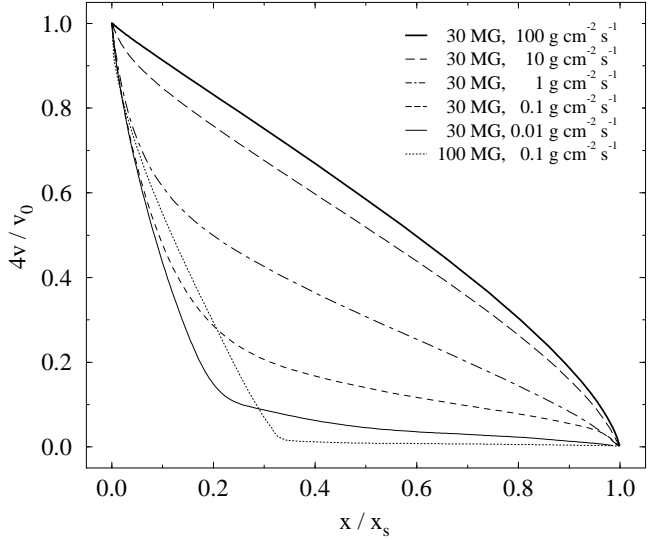}
\caption[]{\label{vnorm} Normalized velocity $w = 4\,v/$\vo\ for \mwd
= 0.6\,\msun\ and the same values of field strength and mass flow
density $\dot m$ as in Fig.~\ref{tnorm}.}
\end{figure}

The initial rise of the individual temperature profiles is similar and
is very rapid following approximately \te\,$ \propto x^{0.35}$
(Fig.\,\ref{tempprof}). One half of \tm\ is reached at 0.001\,\xs\ in
the bremsstrahlung-dominated case and at 0.006\,\xs\ near the
non-hydrodynamic limit. Further downstream the profiles differ
substantially. In the bremsstrahlung-dominated case, the peak electron
temperature is reached quickly, while in the cyclotron-dominated flow
it occurs at the same $x$ at which half of the accretion energy has
been radiated away. The reason is that a temperature gradient is
needed to drive about one half of the radiative flux across the shock
front, while the other half enters the white dwarf atmosphere. {\it In
the plane-parallel geometry}, the optically thick radiative transfer
requires the electron temperature at the shock front to stay below the
peak electron temperature : $T_{\rm e,s} < T_{\rm max} \le T_{\rm
brems,max} \sim 0.5\,T_{\rm i,s}$. This is why we opted to start the
integration with the initial values \tes\,= 0 and \tis\ as given by
Eq.\,(\ref{tshock}). Because of the rapid initial rise in $T(x)$, our
results would have been practically the same had we set \tes\,=
0.5\,\tm.

To facilitate the modeling of specific geometries, we provide the
normalized temperature and density profiles for a sequence of $\dot m,
B$ combinations in Table 1.  We also provide best fits to \tm/\tsh\
and \xs\ as functions of \mdotb.

\begin{table*}[t] 
\caption[]{{\it (a) Top: } Temperature profiles \te/\tm($x$/\xs) for a
white dwarf mass of 0.6\,\msun, field strengths of 10, 30, and
100\,MG, and mass flow densities between 100 and 0.01\,\gcs. {\it
First line:} Analytical solution of Eqs.~(\ref{x-w}) and
(\ref{tbrems}).  {\it Subsequent lines:} Profiles for parameter
combinations $B,\,\dot m$ in MG and \gcs, with $B_7$ in units of
$10^7$\,G. All profiles start at ($x$/\xs,\,\te/\tm) = (0,\,0), end at
(1,\,0), and are normalized to a peak value of unity. {\it (b)~Bottom:
} Same for the normalized velocity $w = 4\,v/$\vo.  These profiles
start at ($x$/\xs,\,$w$) = (0,\,1) and end at (1,\,0).}
\vspace*{-2mm}
\begin{flushleft}
\begin{tabular}{@{}l@{\hspace{1mm}}r@{\hspace{2mm}}l@{\hspace{3mm}}c@{\hspace{2mm}}c@{\hspace{2mm}}c@{\hspace{2mm}}c
@{\hspace{2mm}}c@{\hspace{2mm}}c@{\hspace{2mm}}c@{\hspace{2mm}}c
@{\hspace{2mm}}c@{\hspace{2mm}}c@{\hspace{2mm}}c@{\hspace{2mm}}c
@{\hspace{2mm}}c@{\hspace{2mm}}c@{\hspace{2mm}}c@{\hspace{2mm}}
c@{\hspace{2mm}}c}
\noalign{\smallskip}\hline\noalign{\smallskip} 
$\dot m$      & $B$           & \susi         & 
\multicolumn{14}{c}{$T$/\tm ~vs. ~$x$/\xs}\\[0.5ex] 
              &               && \hspace{-7mm}$x$/\xs = 
$10^{-4}$&$10^{-3}$&0.01&0.02 & 0.05  & 0.10  & 0.20  & 0.30  & 
0.40  & 0.50  & 0.60  & 0.70  & 0.80  & 0.90  & 0.95  & 0.98  &
0.99          \\[0.3ex]
\noalign{\smallskip}\hline\noalign{\smallskip} 
      &       & $\infty$      & 
1.000 & 0.999 & 0.995 & 0.989 & 0.973 & 0.945 & 0.886 & 0.821 & 
0.751 & 0.674 & 0.589 & 0.494 & 0.382 & 0.245 & 0.156 & 0.085 &
0.054 \\
100   &  10   & 100           & 
0.230 & 0.515 & 0.916 & 0.984 & 0.995 & 0.967 & 0.905 & 0.840 & 
0.770 & 0.691 & 0.603 & 0.505 & 0.392 & 0.250 & 0.158 & 0.085 &
0.053 \\
10    &  10   & 10            & 
0.230 & 0.515 & 0.915 & 0.984 & 0.996 & 0.968 & 0.906 & 0.842 & 
0.770 & 0.692 & 0.604 & 0.506 & 0.393 & 0.250 & 0.159 & 0.082 &
0.048 \\
1     &  10   & 1             & 
0.219 & 0.499 & 0.898 & 0.977 & 0.998 & 0.973 & 0.914 & 0.850 & 
0.782 & 0.710 & 0.629 & 0.536 & 0.426 & 0.281 & 0.176 & 0.085 &
0.047 \\
0.1   &  10   & 0.1           & 
0.191 & 0.434 & 0.822 & 0.927 & 0.999 & 0.980 & 0.916 & 0.849 & 
0.779 & 0.706 & 0.622 & 0.522 & 0.400 & 0.239 & 0.127 & 0.047 &
0.022 \\
0.01  &  10   & 0.01          & 
0.175 & 0.395 & 0.743 & 0.861 & 0.981 & 0.991 & 0.860 & 0.743 & 
0.650 & 0.565 & 0.476 & 0.384 & 0.285 & 0.165 & 0.092 & 0.039 &
0.021 \\
100   &  30   & 5.75          & 
0.228 & 0.512 & 0.914 & 0.984 & 0.994 & 0.966 & 0.903 & 0.836 & 
0.766 & 0.686 & 0.598 & 0.499 & 0.388 & 0.247 & 0.153 & 0.080 &
0.046 \\
10    &  30   & 0.58          & 
0.224 & 0.504 & 0.909 & 0.981 & 0.994 & 0.958 & 0.888 & 0.813 & 
0.738 & 0.657 & 0.569 & 0.473 & 0.361 & 0.225 & 0.134 & 0.065 &
0.035 \\
1     &  30   & 0.058         & 
0.209 & 0.468 & 0.865 & 0.958 & 0.997 & 0.937 & 0.815 & 0.717 & 
0.629 & 0.546 & 0.460 & 0.372 & 0.272 & 0.162 & 0.091 & 0.042 &
0.025 \\
0.1   & 30    & 0.0058        & 
0.185 & 0.408 & 0.738 & 0.849 & 0.971 & 0.997 & 0.833 & 0.657 & 
0.548 & 0.468 & 0.401 & 0.336 & 0.274 & 0.197 & 0.145 & 0.085 &
0.045 \\
0.01  &  30   & 0.00058       & 
0.178 & 0.390 & 0.710 & 0.802 & 0.922 & 0.994 & 0.878 & 0.652 & 
0.484 & 0.362 & 0.292 & 0.245 & 0.196 & 0.119 & 0.074 & 0.038 &
0.022 \\
100   &  100  & 0.25          & 
0.222 & 0.497 & 0.898 & 0.979 & 0.992 & 0.942 & 0.849 & 0.766 & 
0.685 & 0.603 & 0.517 & 0.424 & 0.317 & 0.188 & 0.104 & 0.044 &
0.020 \\
10    &  100  & 0.025         & 
0.206 & 0.468 & 0.839 & 0.940 & 0.999 & 0.927 & 0.756 & 0.639 & 
0.541 & 0.450 & 0.367 & 0.289 & 0.201 & 0.103 & 0.043 & 0.011 &
0.006 \\
1     &  100  & 0.0025        & 
0.199 & 0.440 & 0.795 & 0.903 & 0.997 & 0.921 & 0.564 & 0.458 & 
0.375 & 0.302 & 0.237 & 0.193 & 0.129 & 0.058 & 0.027 & 0.013 &
0.007 \\
0.1    &  100  & 0.00025       & 
0.180 & 0.354 & 0.626 & 0.727 & 0.880 & 0.976 & 0.966 & 0.600 & 
0.131 & 0.097 & 0.084 & 0.077 & 0.063 & 0.047 & 0.038 & 0.030 &
0.022 \\
\noalign{\smallskip}\hline\noalign{\smallskip}
\end{tabular}

\begin{tabular}{@{}l@{\hspace{1mm}}r@{\hspace{2mm}}l@{\hspace{3mm}}c@{\hspace{2mm}}c@{\hspace{2mm}}c@{\hspace{2mm}}c
@{\hspace{2mm}}c@{\hspace{2mm}}c@{\hspace{2mm}}c@{\hspace{2mm}}c
@{\hspace{2mm}}c@{\hspace{2mm}}c@{\hspace{2mm}}c@{\hspace{2mm}}c
@{\hspace{2mm}}c@{\hspace{2mm}}c@{\hspace{2mm}}c@{\hspace{2mm}}
c@{\hspace{2mm}}c}
\noalign{\smallskip}\hline\noalign{\smallskip} 
$\dot m$      & $B$           & \susi         & 
\multicolumn{14}{c}{$w = 4v$/\vo ~vs. ~$x$/\xs}\\[0.5ex] 
              &               && \hspace{-7mm}$x$/\xs = $10^{-4}$
    & $10^{-3}$ & 0.01 & 0.02 & 0.05  & 0.10  & 0.20  & 0.30  & 
0.40  & 0.50  & 0.60  & 0.70  & 0.80  & 0.90  & 0.95  & 0.98  &
0.99          \\[0.3ex]
\noalign{\smallskip}\hline\noalign{\smallskip} 
      &       & $\infty$      & 
1.000 & 0.999 & 0.992 & 0.984 & 0.960 & 0.921 & 0.841 & 0.761 & 
0.678 & 0.594 & 0.506 & 0.413 & 0.311 & 0.193 & 0.121 & 0.065 &
0.041 \\
100   &  10   & 100           & 
1.000 & 0.999 & 0.992 & 0.985 & 0.961 & 0.920 & 0.840 & 0.760 & 
0.676 & 0.593 & 0.504 & 0.411 & 0.310 & 0.191 & 0.119 & 0.064 &
0.039 \\
10    &  10   & 10            & 
1.000 & 0.999 & 0.991 & 0.983 & 0.958 & 0.915 & 0.834 & 0.755 & 
0.670 & 0.586 & 0.496 & 0.405 & 0.304 & 0.187 & 0.115 & 0.060 &
0.035 \\
1     &  10   & 1             & 
1.000 & 0.998 & 0.983 & 0.969 & 0.932 & 0.882 & 0.788 & 0.700 & 
0.618 & 0.538 & 0.459 & 0.376 & 0.289 & 0.182 & 0.112 & 0.053 &
0.029 \\
0.1   &  10   & 0.1           & 
1.000 & 0.993 & 0.934 & 0.888 & 0.794 & 0.696 & 0.571 & 0.480 & 
0.404 & 0.336 & 0.279 & 0.218 & 0.158 & 0.089 & 0.046 & 0.017 &
0.008 \\
0.01  &  10   & 0.01          & 
1.000 & 0.989 & 0.892 & 0.811 & 0.635 & 0.440 & 0.266 & 0.193 & 
0.150 & 0.116 & 0.090 & 0.066 & 0.045 & 0.024 & 0.013 & 0.005 &
0.003 \\
100   &  30   & 5.75          & 
1.000 & 0.999 & 0.990 & 0.981 & 0.953 & 0.913 & 0.831 & 0.751 & 
0.667 & 0.584 & 0.497 & 0.402 & 0.303 & 0.186 & 0.115 & 0.059 &
0.033 \\
10    &  30   & 0.58          & 
1.000 & 0.996 & 0.972 & 0.948 & 0.902 & 0.947 & 0.758 & 0.676 & 
0.597 & 0.517 & 0.437 & 0.352 & 0.264 & 0.159 & 0.094 & 0.044 &
0.022 \\
1     &  30   & 0.058         & 
1.000 & 0.990 & 0.914 & 0.852 & 0.726 & 0.614 & 0.499 & 0.425 & 
0.363 & 0.307 & 0.254 & 0.199 & 0.144 & 0.084 & 0.047 & 0.019 &
0.009 \\
0.1   & 30    & 0.0058        & 
1.000 & 0.990 & 0.902 & 0.829 & 0.664 & 0.480 & 0.286 & 0.206 & 
0.167 & 0.140 & 0.117 & 0.097 & 0.078 & 0.055 & 0.040 & 0.021 &
0.011 \\
0.01  &  30   & 0.00058       & 
0.999 & 0.982 & 0.895 & 0.824 & 0.653 & 0.435 & 0.148 & 0.087 & 
0.062 & 0.045 & 0.036 & 0.029 & 0.023 & 0.013 & 0.008 & 0.004 &
0.003 \\
100   &  100  & 0.25          & 
1.000 & 0.995 & 0.948 & 0.909 & 0.835 & 0.760 & 0.660 & 0.579 & 
0.506 & 0.437 & 0.367 & 0.294 & 0.215 & 0.124 & 0.068 & 0.029 &
0.013 \\
10    &  100  & 0.025         & 
1.000 & 0.989 & 0.886 & 0.807 & 0.643 & 0.494 & 0.367 & 0.302 & 
0.251 & 0.206 & 0.167 & 0.128 & 0.089 & 0.045 & 0.019 & 0.005 &
0.002 \\
1     &  100  & 0.0025        & 
1.000 & 0.982 & 0.860 & 0.758 & 0.535 & 0.282 & 0.124 & 0.099 & 
0.081 & 0.065 & 0.051 & 0.040 & 0.028 & 0.013 & 0.006 & 0.002 &
0.001 \\
0.1    &  100  & 0.00025       & 
0.999 & 0.985 & 0.885 & 0.827 & 0.707 & 0.552 & 0.293 & 0.069 & 
0.012 & 0.009 & 0.008 & 0.007 & 0.006 & 0.004 & 0.003 & 0.003 &
0.002 \\
\noalign{\smallskip}\hline\noalign{\smallskip}
\end{tabular}
\end{flushleft}
\vspace*{-4mm}
\end{table*}

\subsection{Velocity profiles for an infinite layer}

For calculations of the bremsstrahlung emission, we need the profiles
of the mass density which varies as $\rho \propto \upsilon ^{-1}$.
Fig.~\ref{vnorm} shows the normalized velocity profiles for \mwd \,=
0.6\,\msun\ and the same $\dot m,B$ combinations as in
Fig.~\ref{tnorm}. In the limit of pure bremsstrahlung cooling, the
velocity profile is indistinguishable from that given by the inversion
of Eq.~(\ref{x-w}). Increased cyclotron cooling causes a similar
depression at intermediate $x$ as seen in the temperature
profiles. Table 1 (bottom) provides the velocity profiles in numerical
form for the same parameters as above.

\subsection{Maximum electron temperature \tm}

\begin{figure}[t]
\includegraphics[angle=270,width=8.7cm]{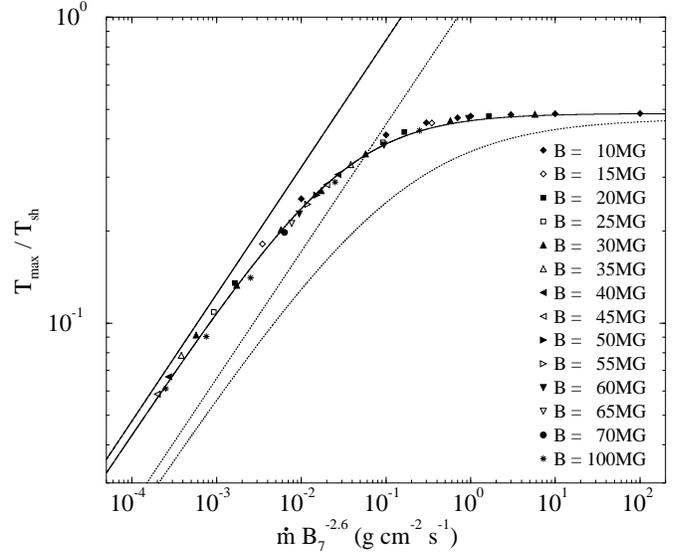}
\caption[]{\label{tm} Maximum electron temperature \tm\ as a function
of \susi\ for \mwd = 0.6\,\msun\ and $B = 10 - 100$\,MG ($B_7 =
B/10^7$\,G).  The solid curve is the fit from Eq.~(\ref{tmtran}), the
solid straight line represents the bombardment solution from
Eq.~(\ref{tmbomb}). The dotted curve and straight line represent the
corresponding fit to the data for 1\,\msun\  (data not shown).}
\end{figure}

In what follows, each model is represented by one ``data point''.
Fig.~\ref{tm} shows \tm/\tsh\ vs. \susi\ for \mwd\,=\,0.6\,\msun\ and
$B = 10 - 100$\,MG.  The dependence of \tm\ on \susi\ is equally well
documented for \mwd\,=\,0.8 and 1.0\,\msun, but for clarity we do not
show these data. The 0.6\,\msun\ results can be fitted by
\begin{equation}
\frac{1}{T_{\rm max}} = \left[\left(\frac{1}{a_0\,T_{\rm
max,bomb}}\right)^\alpha + \left(\frac{1}{a_1\,T_{\rm
max,brems}}\right)^\alpha ~\right]^{1/\alpha}
\label{tmtran}
\end{equation}  
with \tmbo\ from Eq.~(\ref{tmbomb}) and \tmbr\ from
Eq.~(\ref{tmaxbrems}). The exponent $\alpha$ measures the smoothness of the transition
between cyclotron and bremsstrahlung solutions.
The fits for 0.6 and 1\,\msun\ are included in
Fig.~\ref{tm} as the solid and dotted curves, respectively. The fit
parameters $a_0, a_1$, and $\alpha$ are listed in Table 2 for all
three white dwarf masses. The fact that $a_1$ falls slightly short of
1.0 indicates that the limiting value $T_{\rm max} = (\mu/\mu_{\rm
i})T_{\rm i,s}$ for large $\dot m$ is not yet reached at
100\,\gcs. Even at this high $\dot m$, radiative energy losses remove
some energy prior to equipartition. The maximum temperatures at low
$\dot m$ remain about 10\% below the temperatures predicted by the
bombardment solution of WB92,93 (straight lines, $a_0 \simeq 0.9$). A
difference as small as this is actually remarkable considering the
substantially different theoretical and numerical approaches
(radiation hydrodynamics vs. radiative transfer in a static
atmosphere).

\subsection{Column density \xs}

\begin{figure}[t]
\vspace{2mm}
\includegraphics[angle=270,width=8.8cm]{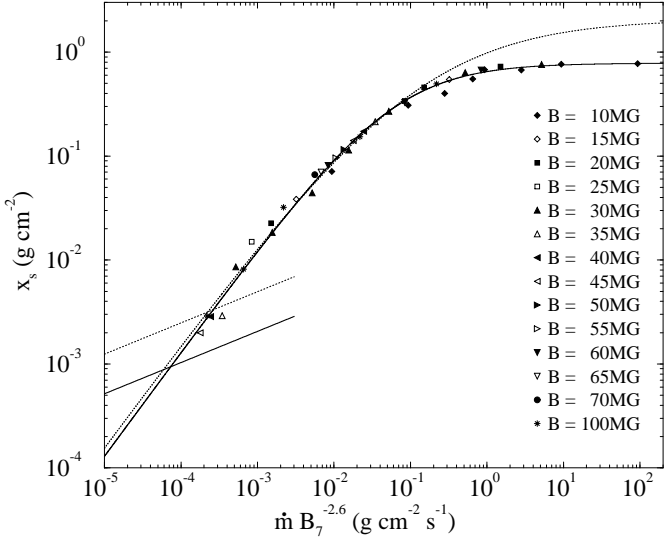}
\caption[]{\label{xs} Column density \xs\ of the post-shock cooling
region as a function of \susi\ for the same parameters as in
Fig.~\ref{tm}. The solid curve is the best fit from
Eq.~(\ref{xstran}). The dotted curve indicates the corresponding fit
to the data for 1\,\msun. The straight lines in the
lower left denote the bombardment solutions from Eq.~(\ref{xsbomb})
for 0.6\,\msun\ and 1.0\,\msun.}
\end{figure}

The transition of \xs\ between the bombardment and the bremsstrahlung
solutions (Eqs. \ref{xsbomb} and \ref{xsbrems}) is more complicated
than that of \tm. Fig.~\ref{xs} shows \xs\ as a function of \susi\ for
\mwd\ = 0.6\,\msun\ and $B = 10 - 100$\,MG. Again, the data points for
0.8 and 1.0\,\msun\ are not shown for clarity. We fit \xs\ by\\[-1ex]
\begin{equation}
\frac{1}{x_{\rm s}} = \left[\left(\frac{\mu}{b_0\,\mu_{\rm e}\,\dot m
B_7^{-2.6}}\right)^\beta + \left(\frac{1}{x_{\rm
s,brems}}\right)^\beta\right]^{1/\beta}
\label{xstran}
\end{equation}  
\vspace{1ex}
with $x_{\rm s,brems}$ from Eq.~(\ref{xsbrems}). The fit parameters
$b_0$ and $\beta$ are given in Table 2 for the three values of
\mwd. Again, the fits are shown for 0.6 and 1.0\,\msun\ (solid and
dotted curve), with the corresponding bombardment solutions added as
straight lines. Note that, contrary to what we found for \tm, the
first term in Eq.~(\ref{xstran}) does not represent the bombardment
solution, but rather the cyclotron-dominated shock heated plasma. It
connects to the bombardment solution as the non-hydrodynamic regime is
approached and bridges a gap of two orders of magnitude in \xs\
between the bombardment and bremsstrahlung solutions. Clearly, the
quantitative determination of $x_{\rm s}(\dot m, B)$ requires
radiation-hydrodynamical calculations.  The $\dot m$-dependence of
Eq.~(\ref{xstran}) reproduces that of Eq.~(\ref{thinxs}), the
molecular weight dependence is added here and taken from
Eq.~(\ref{epscyc}).

\begin{table}[hb]
\caption[ ]{\label{tabtmax} Fit parameters of Eqs.~(\ref{tmtran}),
(\ref{xstran}), and (\ref{hshtran}) for three values of the white
dwarf mass \mwd. }
\begin{flushleft}
\begin{tabular*}{\hsize}{@{\extracolsep{\fill}}cccccccc}
\noalign{\smallskip} \hline \noalign{\smallskip}
$M$  & $a_0$   &$a_1$ & $\alpha$ & $b_0$ & $\beta$ & $c_0$ & $\gamma$ \\
(\msun) &&&&(s)& & (10$^8$\,cm) & \\[0.3ex] 
\noalign{\smallskip} \hline \noalign{\smallskip}
0.6 & 0.91 & 0.968 & 1.67 & 6.5 & 0.70 & 0.95 & 1.0 \\
0.8 & 0.86 & 0.954 & 1.54 & 7.5 & 0.54 & 1.30 & 0.7 \\
1.0 & 0.90 & 0.934 & 1.25 & 8.0 & 0.45 & 1.75 & 0.5 \\
\noalign{\smallskip}\hline\noalign{\smallskip}
\end{tabular*}
\end{flushleft}
\end{table} 

\begin{figure}[t]
\includegraphics[angle=270,width=8.8cm]{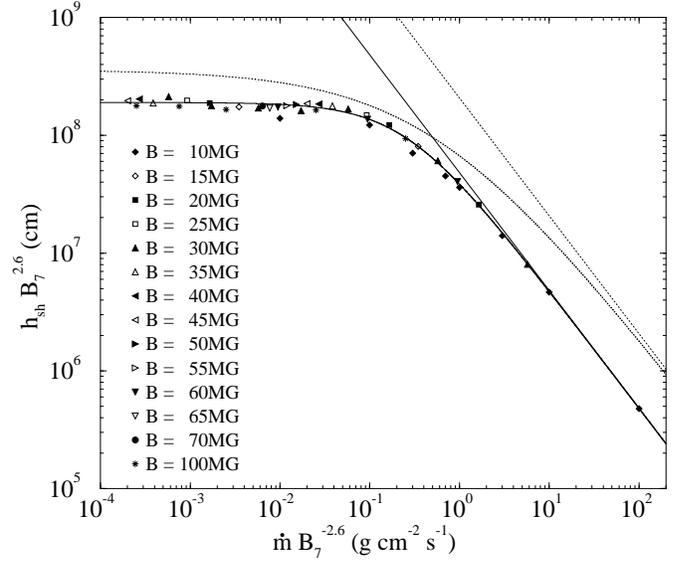}
\caption[]{\label{hshb} Same as Fig.~\ref{xs} but for geometrical
shock height \hsh\ contained in the quantity \hsh $B_7^{2.6}$. The
fits refer to 0.6\,\msun (solid curve) and 1\,\msun (dashed
curve). The straight lines represent the corresponding bremsstrahlung
solutions of Eq.~(\ref{hshbrems}).}
\end{figure}

\subsection{Geometrical shock height \hsh}

Fig.~\ref{hshb} shows the quantity \hsh \,$B_7^{2.6}$ for \mwd\ =
0.6\,\msun\ and for field strengths between 10 and 100\,MG. We fit 
the data by
\begin{equation}
\frac{1}{h_{\rm sh}B_7^{2.6}} = \left[\left(\frac{\mu}{c_0\,\mu_{\rm e}}
\right)^\gamma + \left(\frac{B_7^{-2.6}}{h_{\rm sh,brems}}\right)
^\gamma~\right]^{1/\gamma}
\label{hshtran}
\end{equation}
where $\dot m$ as independent variable enters via $h_{\rm sh,brems}$
from Eq.~(\ref{hshbrems}) and the fit parameters $c_0$ and $\gamma$
are listed in Table 2. The fits for 0.6\,\msun\ and 1.0\,\msun\ are
shown (solid and dotted curve). The limiting dependencies for large
and small \mdotb, respectively, are \hsh$\, = h_{\rm sh,brems} \propto
\dot m^{-1}$ and \hsh = $c_0\,B_7^{-2.6}$ = const., as predicted by
Eq.~(\ref{thinhsh}). The shock height is related to \xs\ by the mean
post-shock density $\overline{\rho}$ = \xs/\hsh\ which is
$\overline{\rho} = 6.97\,\rho_0$ for the bremsstrahlung-dominated
shock solution (see Eqs.~\ref{xsbrems} and~\ref{hshbrems}). For the
cyclotron-dominated shock-heated flow, the mean post-shock density
increases to $\overline{\rho} = (b_0/c_0)\,\rho_0\,\upsilon_{\rm o} 
\simeq 30\,\rho_0$, a result which can not be obtained from simple 
theory.

\begin{figure}[t]
\includegraphics[angle=270,width=8.8cm]{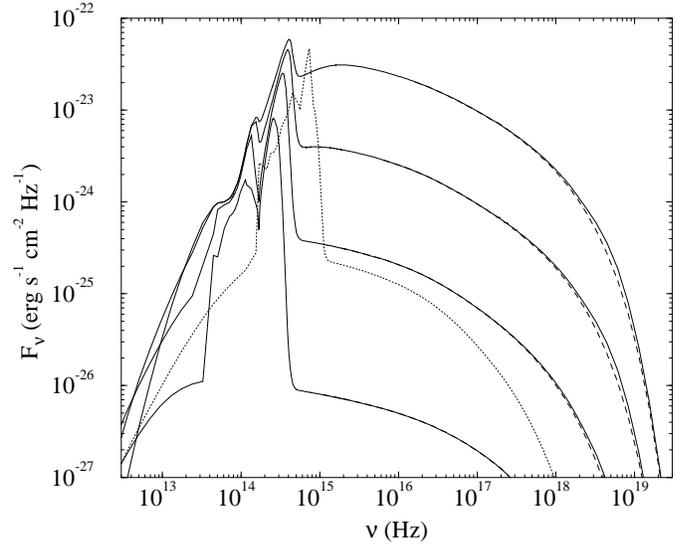}
\caption[]{\label{over06} Overall spectral energy distributions for an
emission region on an 0.6\,\msun\ white dwarf at $d = 10$\,pc with $B
= 30$\,MG, $D = 10^8$\,cm, and $\dot m = 100,10,1$, and
$10^{-1}$\,\gcs\ (from top).  The dashed sections indicate the
emission directed away from the white dwarf at $\vartheta = 5^{\circ}$
without the reflection albedo, the solid curves include the latter.
The dotted curve for $\dot m = 1$\,\gcs, $B = 100$\,MG indicates the
increased cyclotron radiation and the reduced bremsstrahlung flux and
temperature for this field strength.}
\end{figure}

\section{\label{edgeeffects}Emission regions of finite lateral width}

In our radiation-hydrodynamical calculations, energy conservation is
enforced and the radiative luminosity of the infinite layer per unit
area equals the accretion energy $\dot m$\vo. Any real emision region,
however, has a finite lateral width and looses energy not only from
its top and bottom surfaces but also from its sides (Fig.~1). 
Two-dimensional radiation hydrodynamics would then be needed to
calculate the temperature structure. In this section we discuss to
what extent our one-dimensional results are still applicable to
regions of finite extent.

\subsection{\label{spectra}Emitted spectra}

We consider an emission region as depicted in Fig.~1, with finite
width $D$, cross section $D^2$, and field strength $B$. In a first
step, we adopt the temperature and velocity profiles along the flow
lines, $T(x)$ and $\upsilon(x)$, calculated for infinite $D$ also for
the case of finite $D$. 

In the Rybicki code the radiation transport equation was solved with a
mean cyclotron absorption coefficient and electron scattering was
included. For the emission region of finite extent, we calculate the
outgoing flux at angle $\vartheta$ by ray tracing using the
temperature profiles along slanted paths as shown schematically in
Fig.~1. For rays starting or ending on the side surfaces, the
temperature and density profiles were truncated appropriately.  We
account separately for the cyclotron emissivities in the ordinary ray
(index o) and the extraordinary ray (index e), and add 50\% of the
free-free emissivity with Gaunt factor to both. We neglect electron
scattering in the emission region, but include the atmospheric albedo
$A_{\nu}$ (van Teeseling et al. 1994).  Each ray yields a contribution
$\Delta {\cal I}_{\nu}$ to the integrated intensity ${\cal I}_{\nu}$
(in erg\,s$^{-1}$\,Hz$^{-1}$\,sr$^{-1}$) in that direction and the
summation is extended over $n$ rays,
\begin{eqnarray}
\lefteqn{{\cal I}_{\nu}(\vartheta) = \sum_{i=1}^n \Delta {\cal I}_{\nu}
(\vartheta) =  \sum_{i=1}^n 
\,\Delta \sigma^{(i)}\,\Big(\int\limits_0^{s_{\rm max}^{(i)}}\epsilon_{\nu,{\rm o}}^{(i)}
(\vartheta, s)\,{\rm e}^{-\tau_{\nu,{\rm o}}^{(i)}(\vartheta,s)}\,{\rm d}s  }
 \nonumber\\
& & \hspace*{15mm}+~~ 
\,\int\limits_0^{s_{\rm max}^{(i)}}\epsilon_{\nu,{\rm e}}^{(i)}
(\vartheta, s)\,{\rm e}^{-\tau_{\nu,{\rm e}}^{(i)}(\vartheta,s)}\,{\rm d}s 
\Big),
\label{rays}
\end{eqnarray}
where $\tau_{\nu,{\rm o}}^{(i)}(\vartheta,s)$ and $\tau_{\nu,{\rm
e}}^{(i)}(\vartheta,s)$ are the optical depths of the ordinary and
extraordinary rays along path $i$ at angle $\vartheta$, $s_{\rm
max}^{(i)}$ is the pathlength along that ray and $\Delta
\sigma^{(i)}$ the effective projected area associated with it. The
albedo contribution at $\vartheta$ is calculated as $A_{\nu}\,{\cal
I}_{\nu} (\pi-\vartheta)$ and is not yet included in Eq.\,\ref{rays}.
Reprocessing of the flux absorbed in the white dwarf atmosphere is not
considered in this paper and the corresponding flux is, therefore,
missing from our spectra. The spectral luminosity $L_{\nu}$ is
obtained by integrating Eq.~\ref{rays} over $4\pi$ and the total
luminosity $L$ by integration over all frequencies.

\begin{figure}[t]
\includegraphics[angle=270,width=8.8cm]{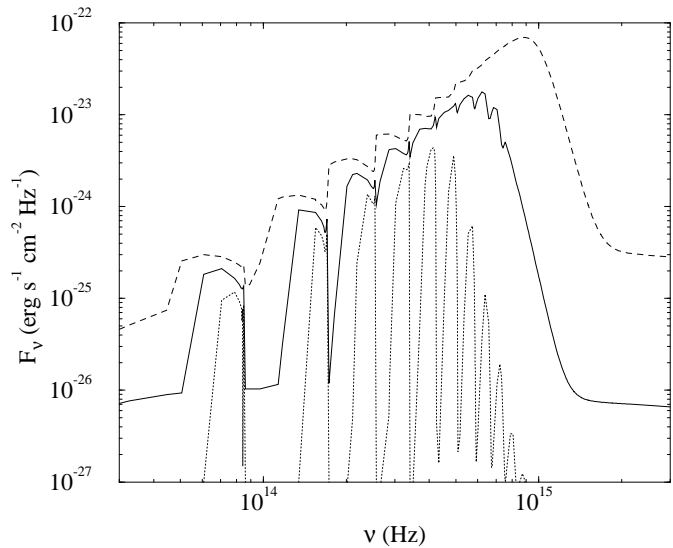}
\caption[]{\label{cyc} Cyclotron section of the spectral energy
distributions for the same set of parameters as in Fig.~8, except
$\vartheta = 80^{\circ}$, and for mass flow densities $\dot m = 1,
10^{-1}$, and $10^{-2}$\,\gcs\ (from top).}
\end{figure}

Fig.~\ref{over06} shows the spectral flux at $\vartheta = 5^\circ$
against the field direction emitted by an emission region with $B =
30$\,MG and an area of $10^{16}$\,cm$^2$ ($D = 10^8$\,cm) on an
0.6\,\msun\ white dwarf at a distance of 10 pc. Cyclotron emission
dominates for low $\dot m$ and bremsstrahlung for high $\dot
m$. Free-free absorption becomes important near $10^{15}$\,Hz at the
highest $\dot m$, but in reality this spectral region is dominated by
the quasi-blackbody component produced by reprocessing of the incident
flux in the white dwarf atmosphere. The results of WB96 on the ratio
of the cyclotron vs. bremsstrahlung luminosities as a function of
$\dot m$ and $B$ remain basically valid, but will be modified if the
shock is buried in the atmosphere and X-ray absorption is accounted
for.

Fig.~\ref{cyc} illustrates the optical depth dependence of the
cyclotron spectra at $\vartheta = 80^{\circ}$. Cyclotron emission
lines at low $\dot m$ change into absorption features at high $\dot
m$. Since in real emission regions the fractional area of the
high-$\dot m$ section is small (Rousseau et al. 1996) observed spectra
show emission lines.

\subsection{\label{luminosity}Specific luminosity}

An isolated emission region of lateral width $D$, shock height \hsh,
and the temperature profile $T(x)$ of the infinite layer appropriate
for the mass flow density $\dot m$ will have ${\cal L} = \dot m$\vo\
for optically thin and ${\cal L} \ge \dot m$\vo\ for optically thick
emission. The overestimate in the latter case results from radiation
emerging from the sides of the region without a compensating
influx. For the optical depths considered here, bremsstrahlung is
practically free of such overestimate, cyclotron radiation is not.

\begin{figure}[t]
\includegraphics[angle=270,width=8.8cm]{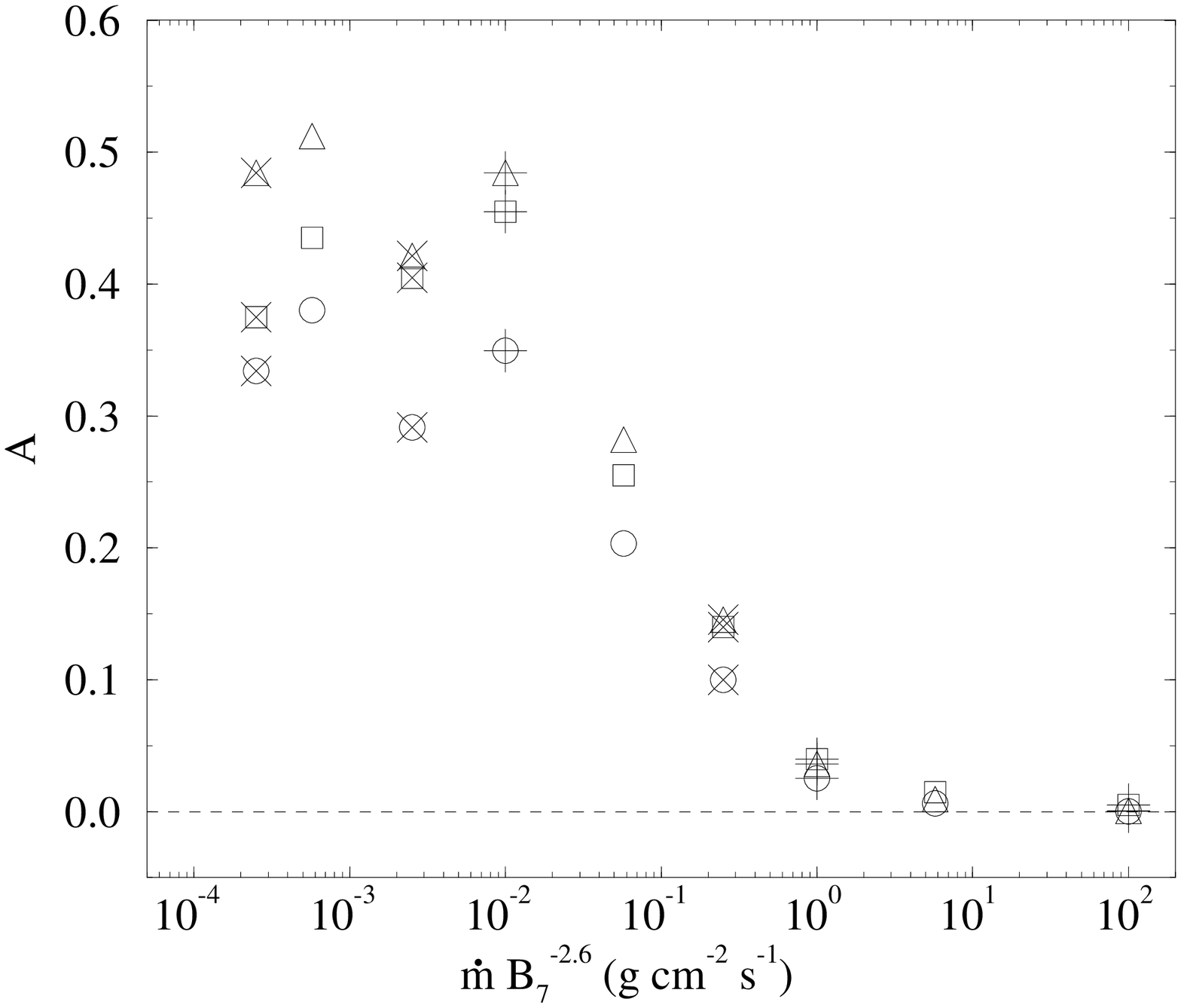}
\caption[]{\label{edge} Quantity $A$ from Eq.~(\ref{ledge2}) measuring
the excess luminosity ${\cal L}/\dot m$\vo\ as function of the aspect
ratio of the emission region, \hsh/$D$. The symbols refer to narrow and
pillbox-shaped emission regions with $D$/\hsh = 0.1
$(\raisebox{-0.3ex}{\pssym{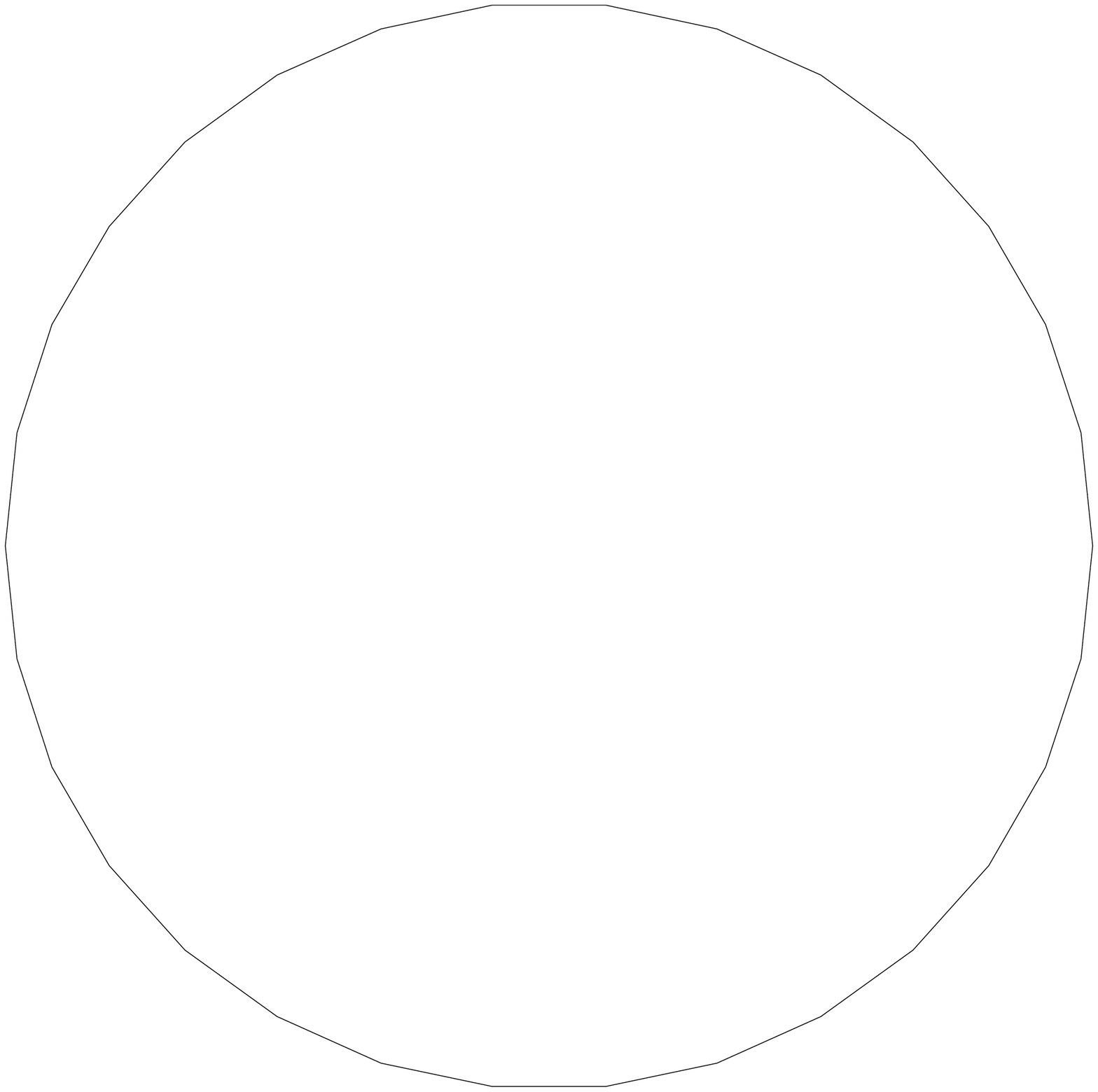}}), D$/\hsh $= 1
(\raisebox{-0.3ex}{\pssym{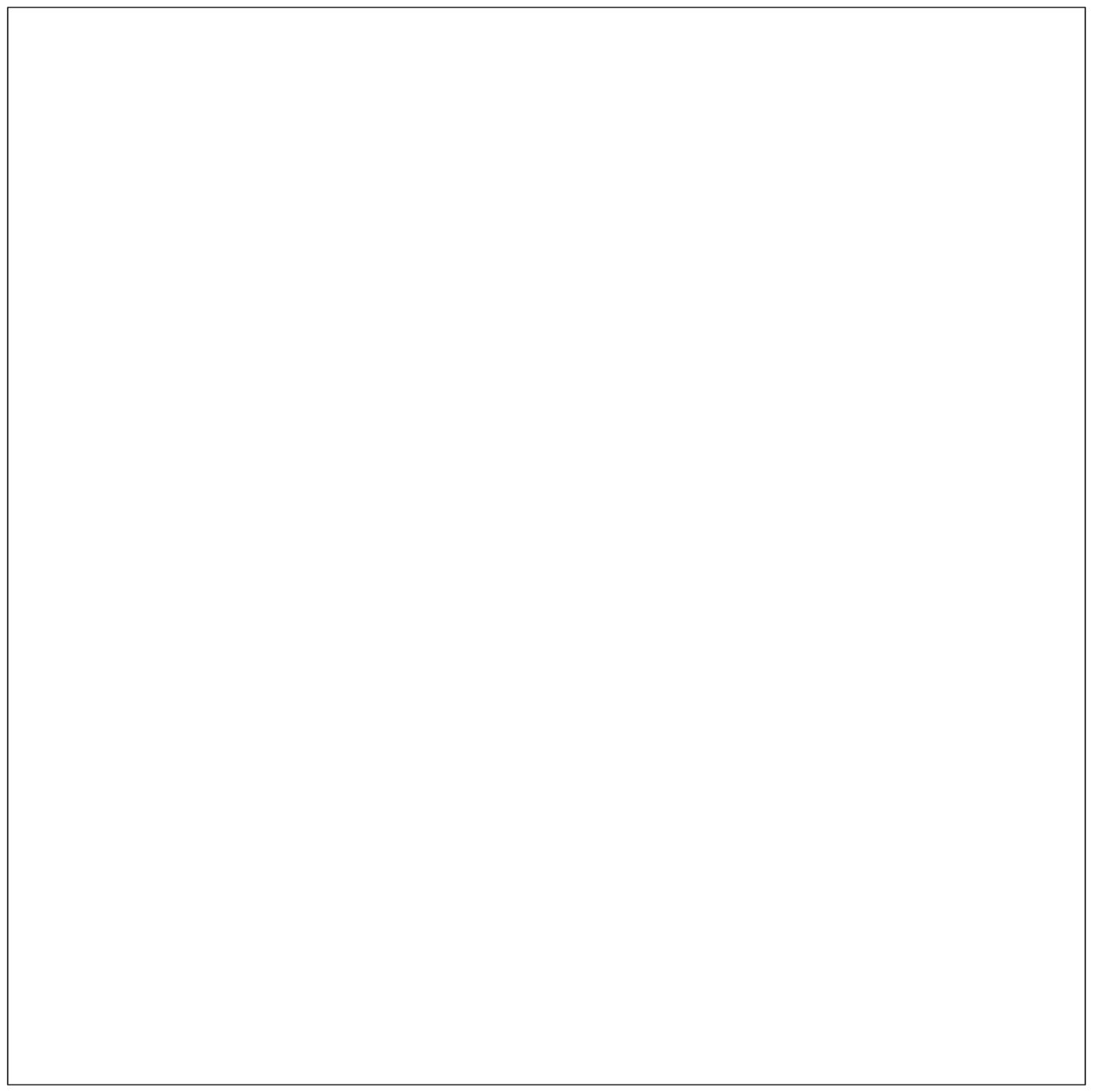}}), D$/\hsh $= 10
~(\raisebox{-0.3ex}{\pssym{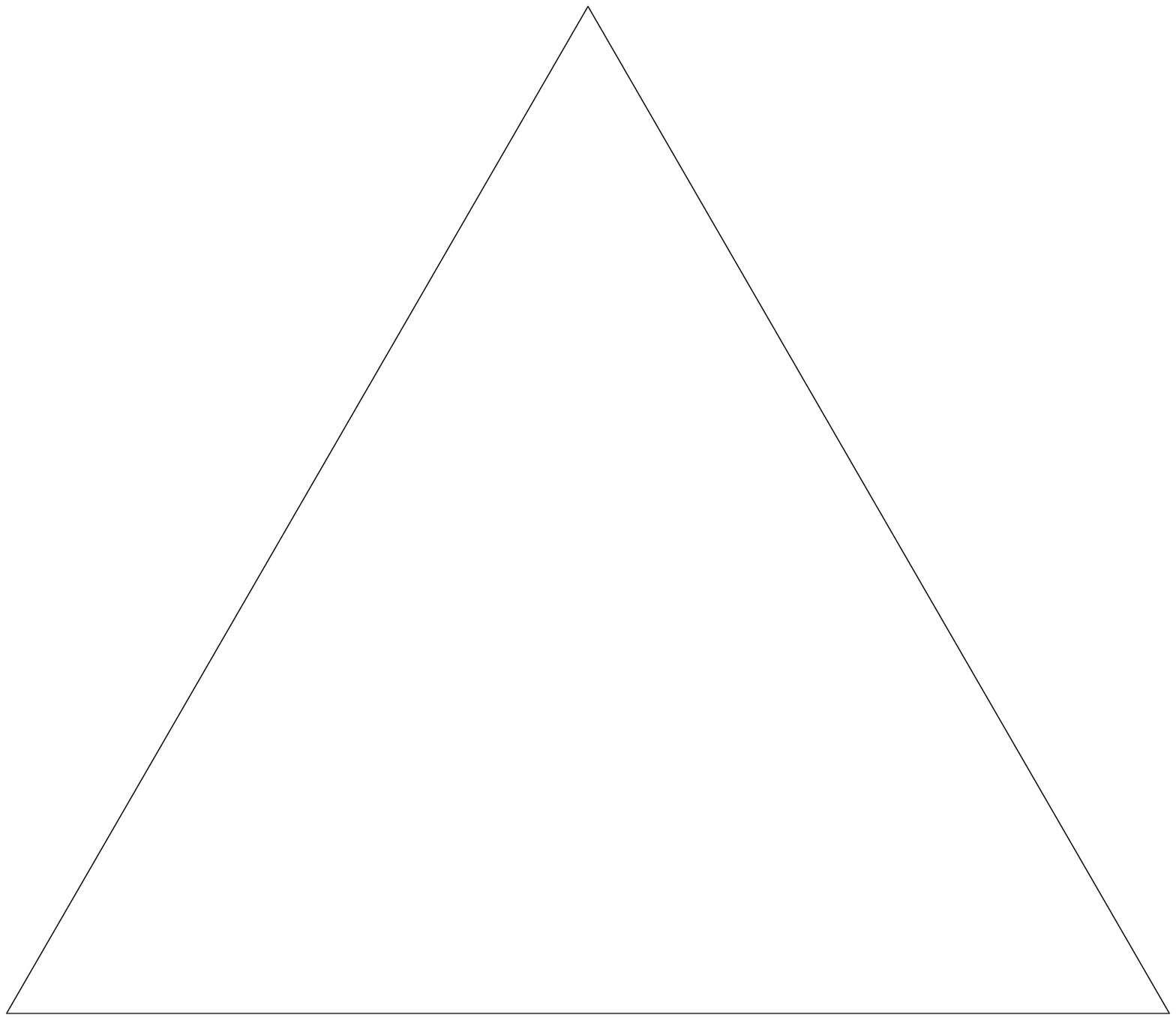}})$, the overplotted + and
$\times$ refer to field strengths of 10 and 100\,MG, respectively, the
uncrossed symbols to $B = 30$\,MG.}
\end{figure}

Let us assume for the infinite layer that $\dot m \upsilon_{\rm o}$
feeds two components of ${\cal L}$, namely ${\cal L}_{\rm thin}$ and
${\cal L}_{\rm thick} = \dot m \upsilon_{\rm o} - {\cal L}_{\rm
thin}$. For finite $D$, we then have
\begin{equation}
{\cal L} \simeq {\cal L}_{\rm thin} + \left(\dot m \upsilon_{\rm o} -
{\cal L}_{\rm thin}\right)\left(1+\frac{2\,h_{\rm sh}}{D}\right) \ge
\dot m \upsilon_{\rm o} ,
\label{ledge1}
\end{equation}
where the second term in brackets is the ratio of the total surface
area of the emission region over the sum of top and bottom areas. For
simplicity, we have neglected the temperature variation over the
surface of the emission region and taken the energy loss per unit area
as constant. We rearrange the terms in Eq.~(\ref{ledge1}) to form a
quantity $A$ which relates ${\cal L}/\dot m \upsilon_{\rm o}$ to the
aspect ratio of the emission region, \hsh/$D$,
\begin{equation}
A = \left(\frac{{\cal L}}{\dot m \upsilon_{\rm o}}-1\right)\frac{D}
{2\,h_{\rm sh}} \simeq 1 - \frac{{\cal L}_{\rm
thin}}{\dot m \upsilon_{\rm o}}.
\label{ledge2}
\end{equation}
Fig.~\ref{edge} shows $A$ as a function of \susi, calculated for model
columns with \hsh/$D = 0.1, 1$ and 10 in the way described in the
previous section. To a first approximation, $A$ is independent of
\hsh/$D$ and the relative luminosity error $({\cal L} - \dot m )/\dot
m \upsilon_{\rm o}$ increases proportional to \hsh/$D$ for a given
\susi.  The quantity $A$ is negligibly small for large \susi\ where
bremsstrahlung dominates and reaches $A \simeq 0.4$ for low \susi\
where cyclotron radiation dominates. Note that $A$ never reaches the
optically thick limit of unity because bremsstrahlung and optically
thin cyclotron emission always contribute.  To give an example,
\hsh/$D$ = 0.5 and $A \simeq 0.4$ imply ${\cal L}/\dot m \upsilon_{\rm
o} \simeq 1.4$, i.e. an overestimate of the luminosity by 40\%.

In order to assess the size of the possible error for AM\,Her stars,
we estimate \hsh/$D$ for a typical accretion rate of $\dot M =
10^{16}$\,g\,s$^{-1}$ as a function of $\dot m$. The linear width of
the emission region is $D \simeq (\dot M/\dot m)^{1/2} \simeq
10^8/\dot m^{1/2}$\,cm with $\dot m$ in \gcs.  For a
bremsstrahlung-dominated flow, Eq.~(\ref{hshbrems}) yields
$h_{\rm sh,brems}/D \simeq 0.48/\dot m^{1/2}$ (for \mwd\ = 0.6\,\msun)
which is less than unity since in this case $\dot m \ga 1$\,\gcs. For
a cyclotron-dominated flow, Eq.~(\ref{hshtran}) correspondingly yields
\mbox{$h_{\rm sh,cyc}/D \simeq 2\dot m^{1/2}/B_7^{2.6}$} which is
again less than unity since now $\dot m \la 1$\,\gcs\ and typically
$B_7 > 1$. Hence, ${\cal L}$ is seriously overestimated only for
isolated narrow subcolumns which are not radiatively shielded by
neighboring fluxtubes.

\subsection{\label{tempcorr}Temperature correction}

Let the application of the unmodified one-dimensional temperature
profile $T(x)$
yield a specific luminosity ${\cal L} = r\,\raisebox{0.4ex}{.}\,\dot m
\upsilon_{\rm o}$ with $r > 1$. We can then take then either: (i)
identify the parameters of this emission region with those appropriate
for the increased mass flow rate $\dot m' = r\,\raisebox{0.4ex}{.}\,\dot m$; 
or (ii) recalculate the emission for a reduced mass flow rate $\dot
m'' = \dot m/r$ and identify temperature and emission of that region
as appropriate for the initial $\dot m$. In case (ii), \tm\ and \xs\
are reduced to \tm($\dot m''$) and \xs($\dot m'')$. This approach
demonstrates that the rising sections of the relations displayed in
Fig.~\ref{tm} (Eq.~\ref{tmtran}) and Fig.~\ref{xs} (Eq.~\ref{xstran})
are further depressed for narrow columns, while the horizontal parts,
where optically thin bremsstrahlung dominates, are not affected. Both
approaches secure energy conservation but can not replace a proper
treatment of the problem. They are not recommended for isolated tall
columns.

\section{Discussion}

We have solved the equations of one-dimensional, two-fluid stationary
radiation hydrodynamics for the shock-heated plasma in the emission
regions on accreting magnetic white dwarfs for a wide range of mass
flow densities $\dot m = \rho_0 \upsilon_{\rm o}$ and field strengths
$B$. For given $B$ and \mwd, the peak electron temperature \tm\ and
the column density \xs\ of the emission region are physically related
to $\dot m$ as the independent variable of the theory. They are no
longer independent variables as in the frequently employed
``constant-$\Lambda$ models''.  It is possible, therefore, to
interpret the observed spectral energy distributions of accreting
magnetic white dwarfs in terms of the distribution of mass flow
densities present in their accretion spots.

We now discuss to what extent the application of these results is
limited by the simplifications made in our calculations.  One major
simplification is the assumption of stationarity which implies that we
neglect the possible occurrence of shock oscillations (Imamura et
al. 1996, Saxton \& Wu 1999, and references therein) and that we can
not treat rapid time variability of $\dot m$.  Since our approach can
accommodate a range of $\dot m$ to occur in neighboring columns, the
emitted spectrum will still approximate the true time-averaged
spectrum if $\dot m$ varies only on time scales exceeding the
post-shock cooling time $\tau_{\rm cool} \simeq 4\,h_{\rm
sh}/\upsilon_{\rm o} \sim 1$\,s.  In the presence of shock
oscillations which have periods of order $\tau_{\rm cool}$, our
results yield a mean temperature and column density which need not
agree with the true time-averaged value if the oscillation is
nonlinear (Imamura \& Wolff 1990).

The assumption of a one-dimensional flow implies that we neglect the
convergence of the polar field lines of the white dwarf.  In the
spirit of this approximation, we have included the acceleration of the
post-shock flow by a constant gravity $g = \,$G\mwd/\rwd$^2$ and
neglected the $r$-dependence of $g$.

The assumption of an infinite layer implies that there is no
temperature gradient perpendicular to the flow. This is no restriction
for bremsstrahlung-dominated flows, but in such gradient is always
established in columns of finite width $D$ by optically thick
radiation components and lowers the mean electron temperature averaged
across the column at any position $x$. We have suggested a simple
first-order correction for the implied overestimate in $T(x)$ which
ensures conservation of energy and provides some remedy for narrow
columns with $h_{\rm sh} \sim D$. For very narrow columns with
\mbox{$h_{\rm sh} \gg D$} or absolutely tall columns with
\mbox{$h_{\rm sh} > 0.1$\,\rwd}, the main radiative energy flow may be
sideways and the approach of Wu et al. (1994) becomes more
appropriate. In summary, our results are valid whenever \mbox{$h_{\rm sh}
\ll$ \rwd~ {\bf and} $h_{\rm sh} \la D$}.

On the positive side, we consider our largely correct treatment of the
two-fluid nature of the post-shock flow. One-fluid treatments
(e.g. Chevalier \& Imamura 1982, Wu et al. 1994) can account for
cooling by cyclotron radiation in addition to bremsstrahlung, but are
limited, by definition, to mass flow densities sufficiently high to
ensure quick equilibration of electron and ion temperatures. They can
not describe the substantial reduction of the peak electron
temperature below the one-fluid value which we show to be present at
low mass-flow density $\dot m$ and/or high magnetic field strength
$B$.  As a result, our description catches the essential properties of
such columns: \mbox{(i) dominant} cyclotron cooling causes the peak
electron temperature to stay far below the peak temperature of the
one-fluid approach; (ii) cyclotron cooling causes a drastic reduction
in the column density and the geometrical shock height of the
post-shock flow compared with pure bremsstrahlung cooling; and (iii)
peak temperature and column density vary smoothly between the two
limiting cases, the bremsstrahlung-dominated high-$\dot m$ regime
(Aizu 1973) and the cyclotron-dominated low-$\dot m$ bombardment
solution (WB92, WB93).  The latter denotes the transition to the
non-hydrodynamic regime and, gratifyingly, our calculations recover
the bombardment solution at the lowest mass flow densities
accessible. Compared with WB96, we obtained numerically more accurate
results and have cast these into simple-to-use fit formulae which
facilitate the modeling of emssion regions within the geometrical
limitations noted above. No other two-fluid calculations with the full
optically thick radiative transfer  are available.

The remaining, mainly geometrical limitations of our \mbox{approach}
are inherently connected to the one-dimensional \mbox{radiative}
transfer. Extension of the calculation to two dimensions encounters
two problems: (i) a substantial increase in complexity and (ii) the
introduction of an additional free parameter in form of the lateral
width of the emission region. Therefore, we consider our
one-dimensional approach with the correction explained in Sect.\,3 as
a reasonable compromise, with the noted exception of tall columns.

The present results can be used to quantitatively model the emission
regions on accreting magnetic white dwarfs.  The discussion of the
overall spectral energy distribution of such objects requires to
account for shocks being buried in the photosphere of the white
dwarf. Such model allows to obtain the $\dot m-$distribution in the
accretion spot from observational data and is presented in Paper II of
this series. The emission properties of AM Herculis binaries depend
not only on $\dot m$ but vary also systematically with field strength:
this dependence is described in Paper III.

\acknowledgements{This work is based on a code originally devised by
U. Woelk. We thank B.T. G\"ansicke, F.V. Hessman and K. Reinsch for
numerous discussions and the referee J. Imamura for helpful comments
which improved the presentation of the results. This work was
supported in part by BMBF/DLR grant 50\,OR\,9903\,6.
}

\end{document}